\documentclass[12pt,preprint]{aastex}

\shorttitle{NGC~1333~IRAS~03256+3055}
\shortauthors{Hodapp, Bally, Eisl\"{o}ffel, Davis}

\slugcomment{Accepted for publication in The Astronomical Journal}

\begin{document}

\title{An S-shaped outflow from IRAS~03256+3055 in NGC~1333}
\author{Klaus W. Hodapp\altaffilmark{1}, 
John Bally\altaffilmark{2},
Jochen Eisl\"{o}ffel\altaffilmark{3},
Christopher J. Davis\altaffilmark{4} }

\altaffiltext{1}{
Institute for Astronomy, University of Hawaii,\\
640 N. Aohoku Place, Hilo, HI 96720,
\\email: hodapp@ifa.hawaii.edu }

\altaffiltext{2}{
University of Colorado\\
Astrophysical \& Planetary Sci., 391 UCB, Boulder, CO 80309,\\
email: bally@origins.colorado.edu}

\altaffiltext{3}{
Th\"{u}ringer Landessternwarte Tautenburg\\
Tautenburg, Germany
\\email: jochen@tls-tautenburg.de }

\altaffiltext{4}{
Joint Astronomy Center\\
660 N. Aohoku Place, Hilo, HI, 96720\\
email: cjdavis@jach.hawaii.edu}

\begin{abstract} 

The IRAS source 03256+3055 
in the NGC~1333 star forming region 
is associated with extended sub-millimeter emission of
complex morphology, showing multiple clumps.
One of these is found to coincide with the driving
source of a bipolar jet of S-shaped morphology
seen in the emission lines of H$\alpha$ and [SII] 
as well as in the H$_2$ emission 
lines in the {\it K}-band.
Detailed images of the driving source 
at the wavelengths of H$\alpha$ and [SII] and
in the {\it I}, {\it J}, {\it H}, and {\it K}-bands 
as well as a {\it K}-band spectrum and polarimetry are discussed.
The near-infrared morphology is characterized by a combination of 
line emission from the jet and scattered light
from a source with a steep continuum spectrum. 
The morphology and proper motion of the jet are discussed
in the context of a binary system with a precessing disk.
We conclude that the molecular core associated with 
IRAS 03256+3055 consists of several clumps, only one of
which shows evidence of recent star formation at optical and
near-infrared wavelengths.
We also briefly discuss a second, newly found near-infrared source associated
with a compact sub-millimeter continuum source near IRAS 03256+3055, and 
conclude that this source may be physically unrelated the cluster of
molecular clumps.

\end{abstract}


\section{Introduction}

Young low-mass stars form by accretion of material through a disk,
a process accompanied by outflow activity.
Objects in their main accretion phase (class 0) are usually associated
with powerful outflows that manifest themselves in 
shock-excited $H_2$ emission.

NGC~1333 is 
one of the youngest regions of clustered 
low-mass star formation and contains a relatively large
population of class 0 sources \citep{lad03} and numerous outflows,
as was pointed out by
\citet{hod95} and \citet{bal96}.
This paper focuses on a region of far-infrared and extended sub-millimeter emission
in the southern part of the NGC~1333 star forming region that was first cataloged as
IRAS~03256+3055.
We assume NGC~1333 to be at the same distance of 316 pc that
\citet{her98} determined for the IC 348 cluster located in
the same Per OB2 molecular cloud complex.

\citet{bal96} discovered
a system of Herbig-Haro objects associated with IRAS~03256+3055,
indicating outflow activity
and thus recent star formation in the region. 
This paper reports new near-infrared images, spectroscopy,
and polarimetry of the young stellar object (YSO) responsible
for this outflow activity. 
Morphologically, the $H_2$ outflow associated with
IRAS 03256+3055 is remarkable for its S-shaped symmetry. 
A number of possible explanations
for deviations from a straight jet were discussed by \citet{eis97}. 
The changes in the IRAS~03256+3055 jet are smooth without discontinuities or kinks.
Jets with this morphology are usually
interpreted as precessing. The precession of the
accretion disk, and therefore of the jet, is caused by the
misalignment of the disk plane and the orbital plane of 
a binary system.

We also present new sub-millimeter images 
at 450$\mu$m and 850$\mu$m 
that confirm the results of \citet{you03} and show a
new compact sub-millimeter object.

\section{Observations and Data Reduction}

\subsection{Infrared Imaging}
The S-shaped molecular hydrogen jet in the area of IRAS 03256+3055
was included in a wide field image of the NGC~1333 region in the
H$_2$ 1--0 S(1) emission line obtained in the nights of July 30 to August 3, 1996
using the QUIRC camera \citep{hod96} at the University of Hawaii (UH) 2.2m telescope. 
The individual integration time was 300 s. Three frames were co-added at the position
of the jet's driving source (Fig.~3, 1996 image).

A deep {\it K}-band image of this region was subsequently obtained with QUIRC
at the UH~2.2m telescope, on the night of February 7, 1998 (UT).
The individual integration time was 120 s per frame and a total of 30 frames were
co-added to form the image shown in Fig.~1.

For proper motion studies, additional images in the H$_2$ S(1) line
were obtained August 1, 2004 with QUIRC at the UH 2.2m telescope.
These observations were carried out with an integration time
of 60 s per frame and 56 such frames were co-added into the final image (Fig.~3, 2004 image). 

The 1996 and 2004 H$_2$ 1--0 S(1) images of the jet do not show any detail beyond what is visible in
the broad-band {\it K}-band image in Fig.~1 and are therefore
not presented in their full extent in a figure. Only the central areas of
these images are shown in Fig.~3.

The coordinates of the flux maximum in the {\it K}-band near the apparent
origin of the S-shaped jet were measured on
a Two Micron All Sky Survey (2MASS) image where this feature is very faintly visible:
03:28:45.3 +31:05:42 (J2000).
To distinguish this near-infrared object from the larger IRAS source
we will refer to the near-infrared ({\it K}-band) manifestation of the 
driving source of the S-shaped outflow as 
NGC1333 J032845.3+310542 (K).

Higher spatial resolution near-infrared images of this central source of the outflow
were obtained in the night of October 20, 2003 with the United
Kingdom Infrared Telescope (UKIRT) using the UKIRT Fast Track Imager (UFTI) \citep{roc03}.
These {\it J}, {\it H}, and {\it K} images are shown as part of Fig.~2.
The total integration time in the {\it H} and {\it K} bands was 18 minutes, while we integrated
for a total of 36 minutes in the {\it J} band.
The filter bandpasses are the Mauna Kea Observatories standard filters as
defined by \citet{tok02}.
The infrared magnitudes of the bright knot of emission visible primarily in the
{\it H} and {\it K} bands were measured using the IRAF ``apphot'' package with an
object aperture diameter of 2.2$\arcsec$ , and a very wide sky annulus 
to avoid the extended, mostly line emission from the rest of the
nebula. Calibration is based on the faint UKIRT standard P247-U.
The knot of continuum emission has the magnitudes 
{\it J}=19.9 (18$\mu$Jy), {\it H}=17.0 (165$\mu$Jy), {\it K}=14.9 (712$\mu$Jy).

{\it K}-band imaging polarimetry was obtained with UFTI and the IRPOL polarimetry
module. IRPOL consists of a rotating halfwave retarder and, internal to UFTI,
a Wollaston prism. A warm focal plane mask isolates two horizontal strips so the
ordinary and extra-ordinary beams from each strip can be projected onto the
array simultaneously. A measurement of the degree and angle of polarization takes
a minimum of two halfwave plate settings.
For our {\it K}-band polarization value for this source,
72 individual exposures of 40 s integration time were used.
Due to the faintness of the object, only an integral
value for the polarization in an aperture of 2.2$\arcsec$ diameter centered on the
flux maximum in the {\it K} band could be obtained.
The polarization of this flux maximum is p = 10.4 \% $\pm$2 \%
at an angle of $\Theta$ = 110$^\circ$ $\pm$ 5$^\circ$.
The polarization vector is indicated in the {\it K}-band
frame of Fig.~2.

\subsection{Archival Images}
We also use optical images 
of the NGC~1333 star forming region
at the wavelengths of H$\alpha$ and [SII] 
as well as in the {\it I} band 
obtained on October 29, 1997 by \citet{bal01} 
at the Kitt Peak National Observatory (KPNO) 4m Mayall telescope, 
and additional images using
the same equipment, but better CCDs, obtained by the same team 
on October 13 and 14, 2001.
Individual integration times were 400s and 600s for the narrow-band filters,
and 60 s for the {\it I}-band, and sets of 5 dithered exposures were combined
into the final archival images.
The images at the wavelength of H$\alpha$ and [SII] are shown in Fig. 1 for comparison
with the {\it K}-band image.
The proper motion of the HH objects was measured by comparison 
of the 1997 and 2001 images and
the resulting proper motion vectors are included in Fig.~1 and listed in Table 1.

\subsection{Astrometry}
The optical H$\alpha$ and [SII] images in Fig.~1 were registered to the
wide-field QUIRC {\it K}-band image using the IRAF tasks ``geomap'' and
``geotran'' on a set of 11 stars common to all images. The registration of
the images has rms residuals of 0.1$\arcsec$ and is sufficiently 
precise to allow detailed discussion of the relative position
of features at different wavelengths. This is particularly
important since hardly any of the outflow features in the
H$\alpha$ and [SII] images coincide with features in
the {\it K} band.

The optical KPNO images, the near-infrared wide-field QUIRC image,
and the UKIRT images in Fig. 2 were all astrometrically referenced to the 
same star near the driving source of the outflow, 
but outside of the field shown in Fig.~2.
The coordinates of this star were measured on the 2MASS image. 

\subsection{Proper Motion of Herbig-Haro Knots}
Two sets of H$_2$ 1--0 S(1) images taken 8 years apart were
used to measure the proper motion of the eastern bow shock. 
The proper motion was measured by first registering the two images
using the IRAF tasks ``geomap'' and ``geotran'' on a set of field stars, 
and then measuring the
proper motion of the bow shock and the motion of a control sample of
stars using the task ``xregister''. The small residual average displacement
of the stars was treated as a residual alignment error between the two
images and was subtracted from the bow shock proper motion.

Individual HH knots are better defined in
H$\alpha$ and [SII] than at infrared wavelengths. 
Therefore the proper motions of the emission knots
outside of the bow shock
were measured by comparison of
KPNO Mayall 4m telescope prime focus images obtained in 1997 and
2001, using the methods described in \citet{bal01}. 

The resulting proper motion vectors are shown in Fig. 1
and are tabulated in Table 1.
In contrast to the result obtained from H$\alpha$ and [SII]
emission, the H$_2$ 1--0 S(1) bow shock has a proper motion direction
parallel to the jet axis. The difference between these measurements may
be caused by the more complex morphology of the excitation in the
H$\alpha$ and [SII] lines where the emission appears composed of individual
knots rather than a smooth bow shock. This point will be discussed in
more detail in section 3.1.

\begin{deluxetable}{cccrrrrrr}
\tabletypesize{\scriptsize}
\tablecaption{Jet Proper Motions}
\tablewidth{0pt}
\tablehead{
\colhead{RA} & \colhead{Dec} & \colhead{Name} & \colhead{H$\alpha$ V} & \colhead{H$\alpha$ PA} & \colhead{[SII] V} & \colhead{[SII] PA} & \colhead{H$_2$ S(1) V} & \colhead{H$_2$ S(1) PA}\\
\colhead{} & \colhead{} & \colhead{} & \colhead{V (kms$^{-1}$)} & \colhead{PA ($^\circ$)} & \colhead{V (kms$^{-1}$)} & \colhead{PA ($^\circ$)} & \colhead{V (kms$^{-1}$)} & \colhead{PA ($^\circ$)}
}
\startdata
03:28:38.8 & +31:06:01 & HH 340 B1 & 122 $\pm$ 36 & 271 & 111 $\pm$ 21 & 281 & - & - \\
03:28:39.3 & +31:05:53 & HH 340 B2 & 196 $\pm$ 21 & 288 & 224 $\pm$ 13 & 268 & - & - \\
03:28:51.5 & +31:05:45 & HH 343 E & 119 $\pm$ 17 & 105 & 148 $\pm$ 13 & 94 & - & - \\
03:28:52.0 & +31:05:42 & HH 343 D & 97 $\pm$ 12 & 104 & 99 $\pm$ 13 & 64 & - & - \\
03:28:52.3 & +31:05:39 & HH 343 D & 29 $\pm$ 35 & 88 & 151 $\pm$ 19 & 91 & - & - \\
03:28:53.0 & +31:05:36 & HH 343 C & 66 $\pm$ 13 & 116 & 118 $\pm$ 14 & 88 & - & - \\
03:28:53.9 & +31:05:26 & HH 343 B & 55 $\pm$ 10 & 156 & 38 $\pm$ 11 & 90 & - & - \\
03:28:54.3 & +31:05:21 & HH 343 A & 81 $\pm$ 07 & 194 & 32 $\pm$ 8 & 192 & 34 & 124 \\
03:28:54.9 & +31:04:43 & HH 350   & 141 $\pm$ 32 & 155 & 102 $\pm$ 70 & 187 & - & - \\
\enddata
\end{deluxetable}

\clearpage

\subsection{Subaru IRCS Spectroscopy}
Long-slit, low-resolution spectroscopy of
NGC~1333~J032845.3+310542~(K)
was obtained using the Infrared Camera and Spectrograph (IRCS) \citep{kob00}
at the Subaru telescope on December 15, 2003 (UT). We used IRCS in low resolution grism
spectroscopy mode in the {\it K} band with a pixel scale of 58~mas/pixel and
0.6$\arcsec$ slit width. 
The slit was oriented along the extended emission at a position angle of 
40$^\circ$ as indicated in Fig. 3. 
The {\it K}-band seeing during the observation was poor, about 1.2\arcsec~FWHM.
Sky frames were taken separately at a position outside the extended
emission associated with IRAS 03256+3055. Atmospheric absorption was
measured by repeatedly observing the A0V star HR 1026, near NGC~1333.

The data were flat-fielded using a continuum lamp spectrum and were sky
subtracted. Residuals of the OH airglow lines after sky subtraction were further reduced
by measuring them on the object frame, outside of the area
affected by object emission lines, and subtracting them
from the full spectral frame. The spectral frame was divided by
the spectrum of the atmospheric absorption standard, and multiplied
by a black body spectrum of T~=~9480~K, the effective temperature
of an A0V star. The resulting spectrum (Fig.~4, lower panel)
is therefore relative F$_\lambda$ units.

Emission
lines of H$_2$ are seen both north east (above) and south west (below) of the 
strong continuum source (Fig. 4 upper panel), 
the emission being more extended in the south west direction.
The spectrum of the continuum source was extracted from the spectral
frame and is presented in the lower panel of Fig.~4. 
No effort was made to subtract
the H$_2$ emission lines from the continuum, since they
show spatial structure along the slit and any attempt to subtract
them would be model dependent.
Aside from the superposed H$_2$ emission lines and Br$\gamma$ emission, the spectrum is
a featureless continuum, rising steeply toward longer wavelengths.
This is the characteristic appearance of a very young star whose
emission is dominated by the continuum emission from hot dust,
with added components of
Br$\gamma$ emission from disk accretion and shock-excited H$_2$ emission
from a jet.

\subsection{Sub-Millimeter Imaging}
We observed NGC~1333 IRAS 03256+3055 with the 
JCMT Sub-Millimeter Common User Bolometer Array (SCUBA) \citep{hol99} 
on January 28, 1998 under exceptionally good atmospheric conditions,
with 450$\mu$m opacity of 0.67 and 850$\mu$m opacity of 0.13.
The SCUBA jiggle-map procedure was used to obtain maps at 450 $\mu$m and
850 $\mu$m (Fig. 5) of the extended emission and multiple sources in
IRAS 03256+3055. Calibration was based on repeated measurements
of Mars as the primary calibrator and CRL 618 as an unresolved
secondary standard source \citep{san94}, for which we adopt the
flux values 11.2~Jy at 450$\mu$m and 4.56~Jy at 850$\mu$m, consistent
with \citet{you03}.
To obtain information about the local but extended flux maxima in
the sub-millimeter emission of IRAS 03256+3055, we performed sky-subtracted aperture
photometry on the maps. We used an object aperture radius of 15" and
a sky annulus between 20$\arcsec$ and 25$\arcsec$ radius and calibrated directly 
against the maps of CRL 618. The results are listed in Table 2.

\begin{deluxetable}{ccccc}
\tabletypesize{\scriptsize}
\tablecaption{Sub-Millimeter Photometry}
\tablewidth{0pt}
\tablehead{
\colhead{Knot} & \colhead{RA} & \colhead{Dec} & \colhead{450$\mu$m} & \colhead{850$\mu$m}\\
\colhead{} & \colhead{} & \colhead{} & \colhead{(mJy)} & \colhead{(mJy)}\\
}
\startdata
A & 03:28:34.5 & +31:07:04 & 756 $\pm$140 & 271 $\pm$12 \\
B & 03:28:39.0 & +31:05:55 & 635 $\pm$140 & 110 $\pm$12 \\
C & 03:28:42.6 & +31:06:14 & 584 $\pm$140 & 139 $\pm$12 \\
D & 03:28:45.1 & +31:05:48 & 552 $\pm$140 & 102 $\pm$12 \\
\enddata
\end{deluxetable}

\clearpage

\section{Results and Discussion}

\subsection{The Jet}
The string of Herbig-Haro objects outlining the curved jet was first
noted on optical images by \citet{bal96}. What we now recognize as the eastern
half of the outflow was called HH~343, with knots labeled A through F
in order from east to west (see Fig. 1). This string of prominent
knots is redshifted. The diffuse nebulosity at the position of
the central source, detected in H$\alpha$ and [SII] was called  HH~340~A,
while the brightest knot in the western outflow lobe was called HH~340~B.
Apparently, the relationship of these objects and the fact that HH~340~A
is associated with
the driving source of the outflow was not noted by \citet{bal96}.

\begin{figure}
\figurenum{1}
\epsscale{0.7}
\plotone{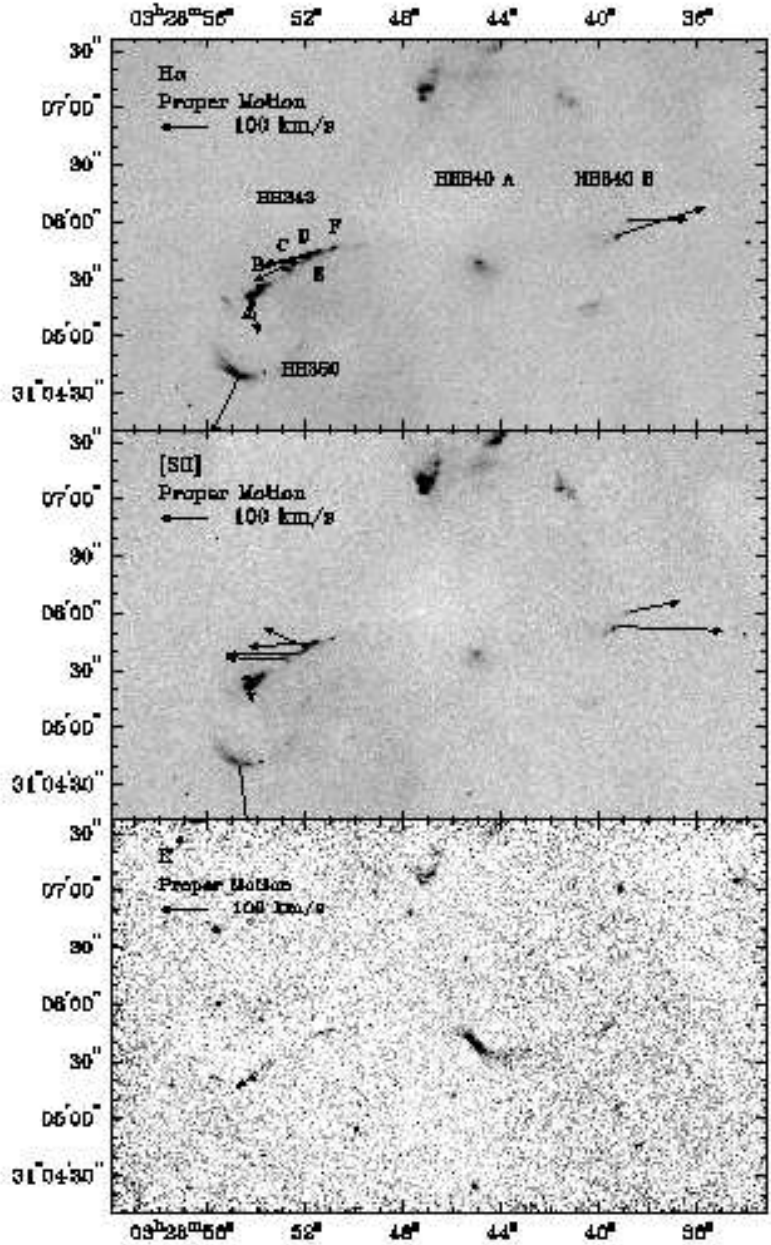}
\caption{
Images of the S-shaped jet associated with IRAS 03256+3055 in NGC~1333 
in H$\alpha$ and [SII] filters, and in the {\it K} band.
The Herbig-Haro objects originally discovered by \citet{bal96} are labeled
in the upper panel. Proper motion vectors are indicated by arrows.
}
\end{figure}
\clearpage

In our infrared images, more extended flux is visible near the central
source and it becomes clear that the HH knots 343 and
340 B form the two lobes of a S-shaped bipolar jet.
Figure 2 shows that the
central driving source 
NGC1333 J032845.3+310542 (K)
is close to but slightly to the north-east of HH~340~A as
seen in H$\alpha$ and [SII].
This diffuse H$\alpha$ and [SII] emission seen near the position of the central source
coincides with the south-western (blueshifted) lobe, as
it is expected in a bipolar structure where the receding (north-eastern) lobe is
more obscured by dust. 
We will discuss the driving source of the outflow in detail in section 3.2.

The {\it K}-band morphology of the jet is very different from that at
the wavelengths of H$\alpha$ and [SII] (Fig. 1). 
The H$\alpha$ and [SII] lines trace much higher excitation
levels than the H$_2$ emission lines that dominate the
emission in the {\it K} band. 
The H$\alpha$ and [SII] images show well defined knots that
led to the original naming HH~343~A-F, while on the {\it K}-band
image, the distinction between the individual knots is not discernable.
Rather, the molecular hydrogen emission consists of lines of enhanced
emission that outline the outflow cavity. 
Similarly, the western lobe shows two distinct knots in H$\alpha$ and [SII]
while the {\it K}-band image shows a contiguous emission region 
that outlines an outflow cavity.

The eastern jet ends in a typical bow shock.
The images in H$\alpha$ and [SII] show two maxima of flux, separated
in an east-west direction, and connected by fainter extended flux. 
One maximum is located east (ahead) of the bow shock traced in
the lower excitation {\it K}-band image dominated by H$_2$ lines, 
while the other maximum is
behind the bow shock. This morphology matches the predictions of
shock models. The highest excitations are found at the apex of the
shock, and in the jet shock (the Mach disk). Lower excitations are
found in the bow shock wings formed by material flowing sideways away from
the main shock region. Other examples of this morphology have been
discussed by \citet{smi03}.

The diffuse emission in the western jet near the central source 
coincides with a ridge of sub-millimeter emission (Fig. 5, lower panel).
The position of the optical knots HH 340 B and the corresponding 
H$_2$ emission coincides with the edge of sub-millimeter clump B.
This does not prove, but suggests that the western jet interacts with
molecular clump B, and may play a role in triggering future star formation
there.

The proper motion vectors are consistent with a curved jet for the system
of Herbig-Haro knots in HH~343 and HH~340.
Approximating the jet by an arc of a circle and dividing by the
proper motion of the bow shock in the near-infrared, a kinematic age
of the jet of $\approx$6000 years is obtained, which represents a lower
limit to the true age of the jet. The projected
jet axis has precessed by almost 90$^\circ$ in this time interval.

The more extended 
bow shock HH~350 \citep{bal96} lies to the south of the tip of HH~343~A, and is seen
in H$\alpha$ and [SII], but is not detected in the {\it K} band. Both its
large size and lack of infrared emission suggest that it is not driven
by the same source as HH~343 and 340. Rather, HH~350 is more likely part
of the extended larger system of shock fronts possibly originating
from IRAS~2 and containing HH~350, HH~14, and HH~351 \citep{bal96}.

While a number of effects can cause an outflow to deviate from a straight
line \citep{eis97}, a jet with S-shaped morphology and
smooth curvature without discontinuities or kinks is most likely the
result of a precessing driving source.
Precessing jets have been discussed by \citet{ter98,ter99}, and \citet{bat00} in the context
of disks in binary systems that are misaligned with the orbital
plane. \citet{bat00} conclude that, typically, the disk will be forced into
alignment with the orbital plane in approximately one precession
period, which in turn is an order of magnitude longer (by about
a factor of 20) than the binary orbital period.
This view is supported by the fact that, to date, no precessing jets with
multiple turns have been observed. The object discussed here is no
exception. From the central driving source to the main bow shock,
presumably the end of the jet, it has precessed by about one quarter turn.
The kinematic age of
$\approx$6000~years provides an estimate for the 
jet's age under the assumption that this bow shock is the surface
of first interaction with previously undisturbed molecular
cloud material. The lack of any indication of other shock fronts farther
away from the driving source makes this a plausible assumption.

Having observed roughly a quarter turn of the jet in $\approx$6000~years gives an estimate
of the precession period of 24000~years, and therefore an estimate of the
disk alignment time of the order of 2$\times$10$^4$~years, i. e., of order of the 
main accretion time on the protostar. Following \citet{bat00} a
binary period of the order of $\approx$1000~years is implied, corresponding to a binary
separation of $\approx$100~AU for a total mass of $\approx$1 M$_\odot$ of the two components,
a reasonable upper limit given the overall faintness of this source.
For any plausible total mass, the binary orbit major axis is of the
order of the size of our own Solar System. At the distance of 316 pc
assumed here for NGC~1333, this corresponds to angular separations of
the order of 0.1\arcsec to 0.3\arcsec. Direct detection of the
two components in the sub-millimeter requires better spatial resolution than
provided by existing interferometers.

\subsection{The Driving Source of the Outflow}
Figure 2 shows detailed images of 
NGC1333 J032845.3+310542 (K)
at wavelengths from H$\alpha$ to the {\it K} band and illustrates the
dramatic changes in morphology with wavelength.
The images at short wavelengths (H$\alpha$, [SII], {\it I}) show an object of
cometary morphology with an apex in the north-east.
At longer wavelengths, in the {\it J}, {\it H}, and {\it K} bands
a northern lobe and a relatively concentrated flux maximum begins to be visible. 
The broad flux minimum that separates the north-eastern and south-western
lobe is most pronounced in the {\it J} and {\it H} bands, but is less
visible in the {\it K} band. This is the typical appearance of a
bipolar nebula where direct radiation from the central star is blocked
by an extended, possibly flared disk seen close to edge-on.
The disk plane in NGC1333 J032845.3+310542 (K) is not precisely oriented
edge-on to the observer, however.
The south-western lobe is pointing toward the observer, consistent
with the redshift found by \citet{bal96} in the HH 343 knots, and is
therefore the only lobe visible at optical wavelengths.
Due to the inclination of the disk we begin to
see scattered light and possibly some direct light from the central star
(or stars) in the {\it H} and {\it K} bands.

\begin{figure}
\figurenum{2}
\epsscale{0.7}
\plotone{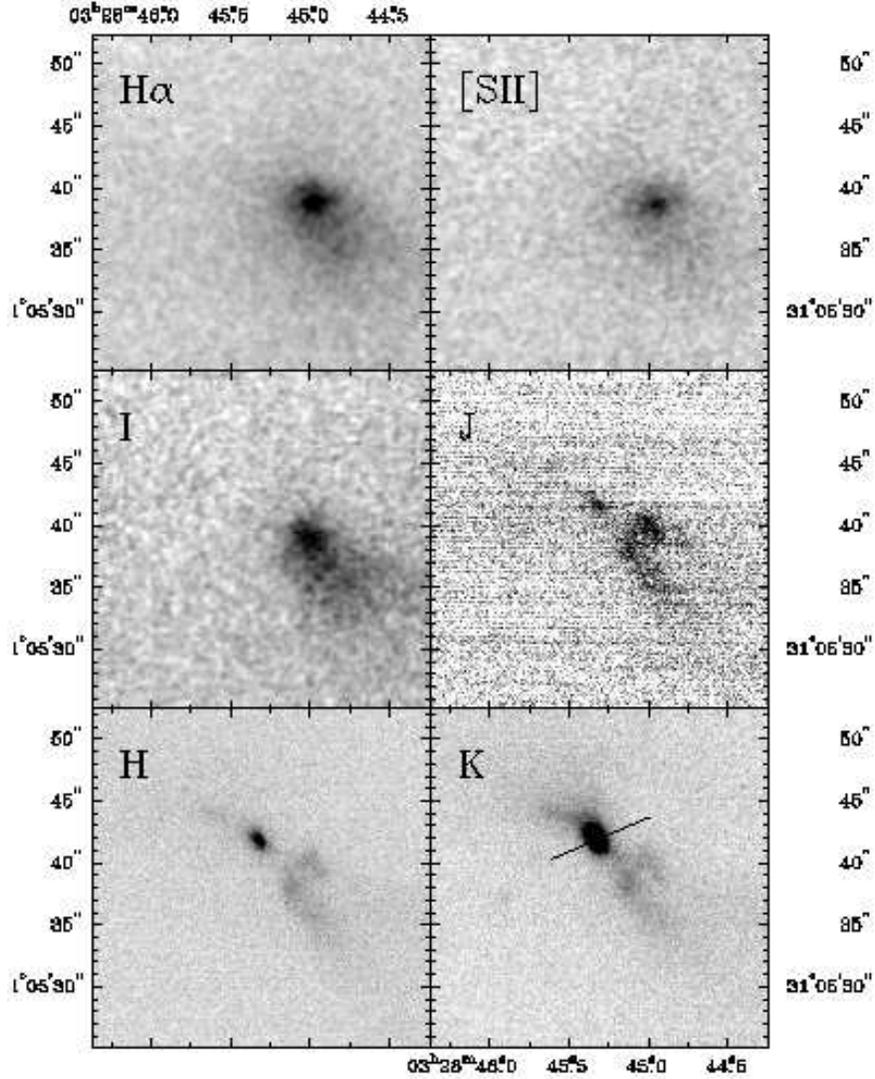}
\caption{
H$\alpha$, [SII], {\it I}, {\it J}, {\it H}, and {\it K}-band images of
NGC1333 J032845.3+310542 (K)
and the nebulosity associated with it.
In the {\it K}-band image, the polarization vector integrated over
the bright emission knot (p=10.4\%, $\Theta$=110$^\circ$) is indicated.
}
\end{figure}
\clearpage

The flux maximum in the {\it K}-band has a
polarization degree of p = 10.4\% $\pm$ 2\%
at an angle of $\Theta$ = 110$^\circ$ $\pm$ 5$^\circ$.
The polarization angle is indicated in the {\it K}-band
panel of Fig. 2.
Its orientation perpendicular to the outflow axis is 
consistent with a strong component of scattered light 
contributing to the emission in this region, and is
typical of deeply embedded class I outflow sources with bipolar morphology \citep{hod84}.

\begin{figure}
\figurenum{3}
\epsscale{0.7}
\plotone{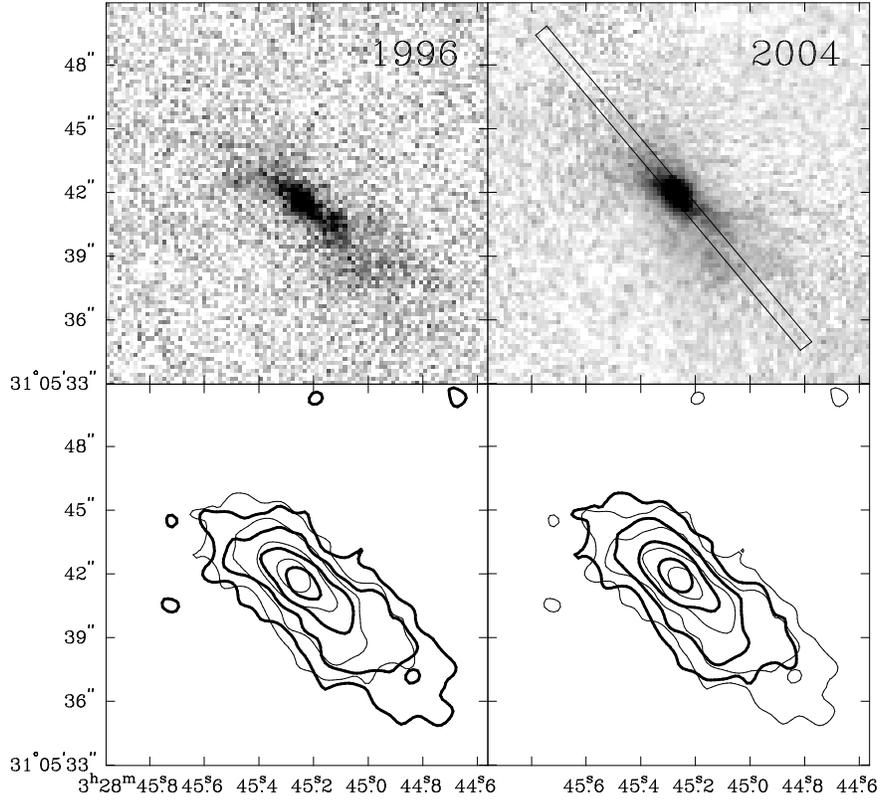}
\caption{
Images of 
NGC1333 J032845.3+310542 (K)
in the $H_2$ S(1) filter obtained in
August 1996 and August 2004, showing the changes in the morphology
of the S(1) line emission and the scattered continuum component
of the bipolar nebulosity associated with this object.
In the contour maps the thick contours correspond to the image
above while the thin contours represent the image at the other
epoch for comparison. The flux was more extended to the south-west in 1996
and the flux peak has shifted by $\approx$0.5$\arcsec$ to 
the north-east between 1996 and 2004.
The orientation and size of the Subaru IRCS slit is indicated
on the 2004 frame.
}
\end{figure}
\clearpage

The morphology of NGC1333 J032845.3+310542 (K)
has changed over the
8 years since we obtained the first infrared image.
Fig. 3 shows images in the H$_2$ 1-0 S(1) emission line and adjacent continuum,
taken 8 years apart with the same telescope, instrument, and filter.
The flux maximum that spectroscopy and polarimetry shows to be scattered
continuum radiation is more prominent in the 2004 image and has shifted by
$\approx$0.5$\arcsec$ to the north-east. Emission
to the south of the brightest knot has diminished over these
8 years.

\begin{figure}
\figurenum{4}
\epsscale{0.7}
\plotone{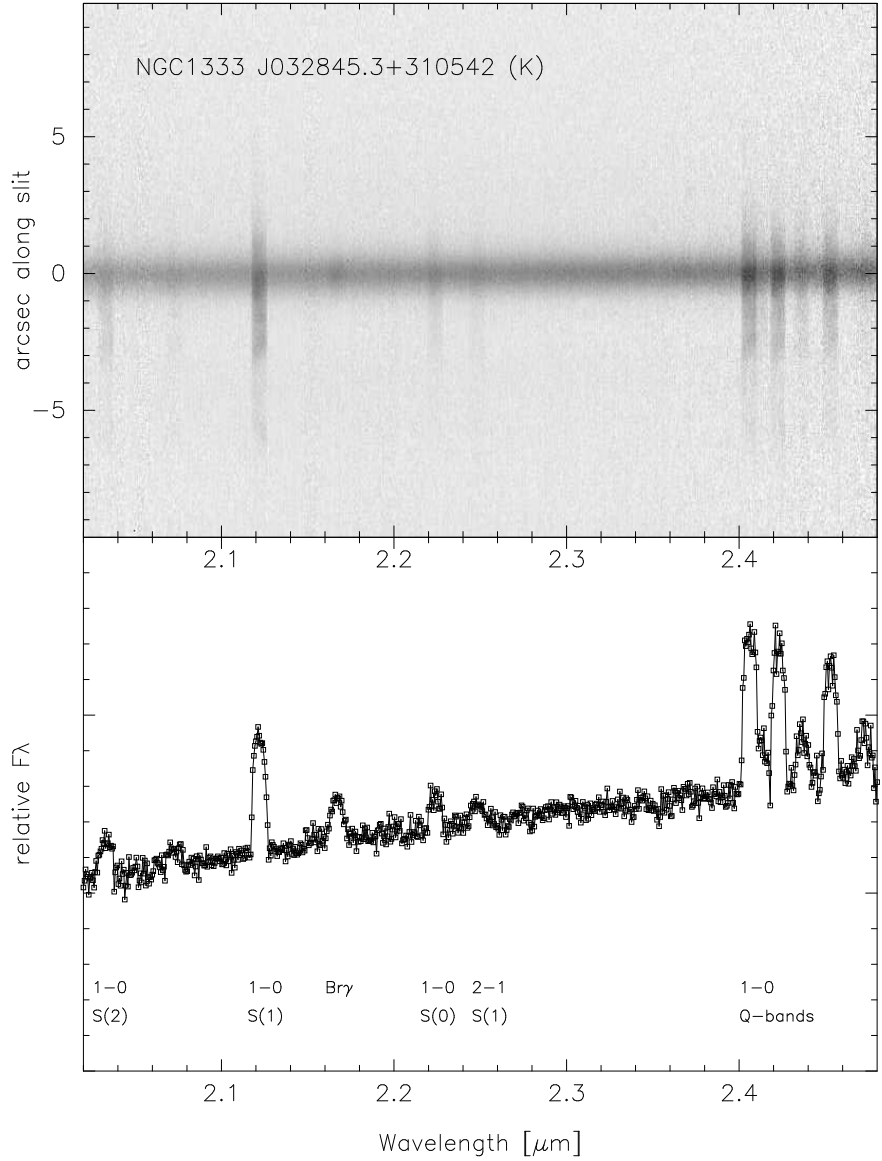}
\caption{
K-band spectrum of 
NGC1333 J032845.3+310542 (K)
and its surrounding nebulosity
taken with the IRCS on the Subaru Telescope. The pixel scale is 58 mas pixel$^{-1}$ 
and the slit was 0.6\arcsec wide, giving a resolving power of $\approx$350.
Slit orientation was
40$^\circ$ as indicated in Fig.~3. 
}
\end{figure}
\clearpage

The {\it K}-band spectrum (Fig.~4) shows continuum emission only from
the bright knot, and shows that the other features
in the extended nebula are line emission from molecular hydrogen.
While we do not have spectra of the object at wavelengths other than
the {\it K} band, we suspect that most of the flux seen in the extended nebula
at shorter wavelengths is also line emission.
It is therefore likely that the driving source of the outflow
is located within the knot of continuum emission at 03:28:45.3 +31:05:42 (J2000)
that dominates the {\it K}-band image.

The {\it K}-band spectrum of 
NGC1333 J032845.3+310542 (K)
is typical for a SED class I source.
$H_2$ emission associated with the jet
is superposed on the continuum spectrum of the central source.
There are no
detectable atomic or molecular absorption
features, but emission of Br$\gamma$ is observed. 
In low resolution spectra of late type stars, including low mass
class II and class III YSOs, the NaI doublet at
2.206$\mu$m and 2.209$\mu$m and the CaI triplet at 2.261$\mu$m, 2.263$\mu$m, and 2.266$\mu$m
should be detectable in addition to the most prominent spectral feature in
such stars, the CO bandheads \citep{hod93}. In their sample of spectra from
stars in the L1641 cluster, the deeply embedded stars have pure continuum
spectra, rising toward longer wavelengths. The most deeply embedded star
in their sample shows H$_2$ hydrogen emission lines in the same way as
NGC1333 J032845.3+310542 (K) does. In the larger study by \citet{gre96} it was 
confirmed that featureless, pure continuum near-infrared spectra are 
usually found in accreting protostars of SED class I, and are far less
common in later SED classes. H$_2$ emission is mostly associated with
class I objects, where it is observed at the base of jets
\citep{dav01}, and is much less prevalent in class II and III.
Note that in our spectrum the line emission is extended along
the IRCS slit, which is aligned with the outflow direction at the source.

In the sample studied by \citet{gre96}, Br$\gamma$ emission was found in
class I and class II objects. The Br$\gamma$ emission in 
NGC1333 J032845.3+310542 (K)
is more localized to the continuum source than the H$_2$ lines. 
Br$\gamma$ emission is in general more likely associated 
with the accretion than the outflow, as
discussed by \citet{muz98} and \citet{dav01}.
It is consistent with our object being a class I source, but does not
in itself strengthen this conclusion.

\subsection{Sub-Millimeter Emission and Evolutionary Status}
Fig. 5 shows the SCUBA map of 
IRAS 03256+3055
and its surroundings.
As was noted by \citet{you03} this object
is an extended sub-millimeter source consisting of several clumps.
The position of the near-infrared flux peak closely coincides 
with clump D (Fig.~4) of sub-millimeter emission farthest to the south east of
the complex emission associated with IRAS 03256+3055.
This clump is
clearly extended compared to the sub-millimeter point source A 
(see section 4) in our images. 
Extended sub-millimeter emission from star forming molecular clumps 
is characteristic of objects still surrounded
by a substantial mass of dust and gas, i. e. class 0 sources \citep{and93}. 

\begin{figure}
\figurenum{5}
\epsscale{0.6}
\plotone{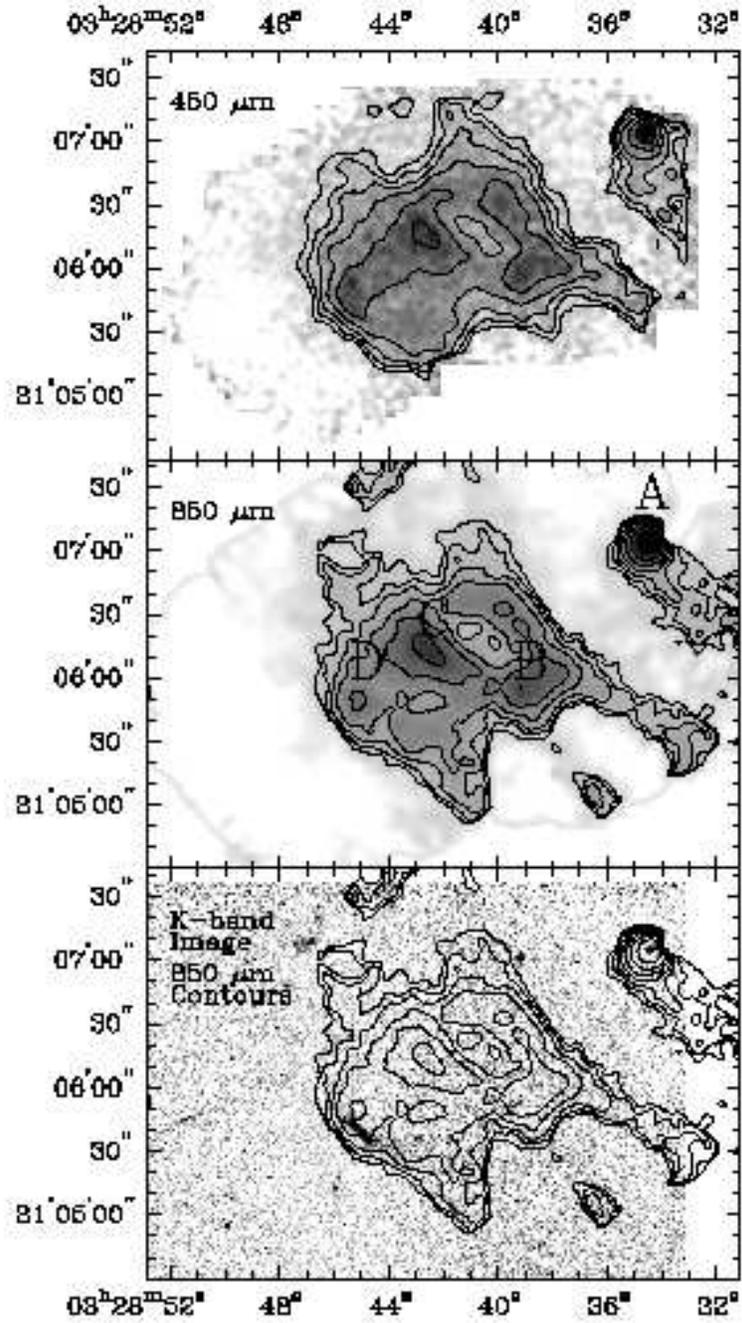}
\caption{
Images of IRAS 03256+3055 in NGC~1333 at 450$\mu$m (upper panel) 
and 850$\mu$m (center panel). The bottom panel shows the superposition
on 850$\mu$m contours on the {\it K}-band image also shown in Fig. 1.
}
\end{figure}

The bolometric temperature, integrated over all
three clumps and the extended emission, is
16 K \citep{you03}, placing this object into the SED class 0.
This value, however, is only based on 60$\mu$m IRAS,
and 450$\mu$m and 850$\mu$m SCUBA data and is therefore
quite uncertain. They estimated the total luminosity to
be only 0.7L$_\odot$ and the total envelope mass to be 
$\approx$ 1.7 M$_\odot$.
The molecular core associated with IRAS 03256+3055
is fragmented into three individual clumps (B, C, and D in
Fig. 5) and some extended emission. The molecular mass of
clump D, which coincides with the driving source of
the outflow, is therefore estimated to be about 1/3 of the
total envelope mass estimated by \citet{you03}.

Additional far-infrared fluxes of IRAS 03256+3055 have been
measured by \citet{cla91} from IRAS co-added scans. He reports fluxes
of 0.02 Jy at 12 $\mu$m, 0.15 Jy at 25 $\mu$m, 1.6 Jy at 60 $\mu$m,
and 6 Jy at 100 $\mu$m. The position of the IRAS source is consistent
with the flux distribution seen in the sub-millimeter maps, i. e., the 
IRAS flux is not concentrated on the near-infrared source.
At the resolution of the IRAS data no extension of the 
object is detectable.
These additional data, showing some mid-IR flux from the
object, point toward a slightly more advanced
state of evolution than class 0. However, the IRAS
data combine the flux from 
NGC1333 J032845.3+310542 (K)
with the other near-infrared source
NGC1333 J032834.5+310705 (K),
that will be discussed below in section 4.
This source 
is most likely a class I source and may have contaminated the low resolution
IRAS data with mid-infrared flux.

The near-infrared spectrum of
NGC1333 J032845.3+310542 (K)
closely resembles that of an object of SED class I.
Therefore, since class 0 and I sources in nearby star forming regions
are readily detectable at sub-millimeter wavelength,
we conclude that
NGC1333 J032845.3+310542 (K)
is the counterpart of the extended clump D of
sub-millimeter emission, and not just a chance superposition.
The well-defined H$_2$ jet with a
young kinematic age and the very low
bolometric temperature 
point to possibly
an even earlier evolutionary state.
The low total luminosity of this region 
indicates a very low mass star or multiple
system at the center of the outflow.
The precession of the jet is best explained by a binary
system with at least one accretion disk.
With the caveat that better data, in particular
higher spatial resolution FIR data, are needed to
firm up this conclusion, we suggest that
NGC1333 J032845.3+310542 (K)
is a low mass binary system of protostars in or near
the end of their main accretion activity phase.

Based on their sub-millimeter observations of the central part of
NGC~1333 \citet{kne00} and \citet{san01} concluded that secondary
star formation in NGC~1333 is triggered in dense shells of gas
swept up by a previous generation of outflows. 
The powerful outflows associated with the well known class 0 and I
sources IRAS 1 - 9 in NGC~1333 \citep{bal96}, all located $\approx$ 1 pc
north of 
the IRAS 03256+3055 group,
may have collectively triggered the
formation of these dense molecular clumps. 
NGC1333 J032845.3+310542 (K)
is the first (binary) protostar in IRAS 03256+3055 to reach 
the class 0 outflow stage, where it develops
a jet readily detectable in the optical and near infrared. 
The jet from NGC1333 J032845.3+310542 (K) appears to impinge on
molecular clump B (Fig.~5), suggesting this may be the next step in
jet-induced triggered star formation in this region.

The IRAS 03256+3055 group of molecular clumps is not
unique. In their extensive studies of the molecular cores in the
$\rho$ Ophiuchus cloud, \citet{mot98} find that most cores 
contain typically of order 10 dense molecular clumps. The typical
linear extent of cores, and the projected separation of the
individual clumps, are similar to that found in the IRAS 03256+3055 group.
Due to the proximity of $\rho$ Ophiuchus (about half the distance
of NGC~1333) and better resolution of their data, they are able
to identify a larger number of individual clumps in each core.
They also find that in the core Oph-A, star formation appears
to have progressed from the outside in, presumably in response
to an external trigger. 
Similarly the cluster of protostellar sources in Serpens NW was studied
by \citet{wil00}. They identify 7 continuum millimeter emission
sources in an area of roughly 0.2 pc extent, some of which are
associated with outflow sources and therefore are protostars,
and some that are probably pre-stellar.

\section{The newly identified YSO NGC1333 J032834.5+310705 (K)}

On our sub-millimeter maps, we found a compact, fairly bright sub-millimeter
source west of the extended emission (labeled A in Fig. 5). 
This source was not
included in the field recorded by \citet{you03}. The compact component of this 
source has a flux of 650 mJy at 450$\mu$m and 274 mJy at 850$\mu$m, measured
by comparing the peak flux in the map to maps of the calibration standard
CRL618.
The compact source is surrounded by extended emission that is mostly visible
at 450 $\mu$m, while it is not clearly distinguishable from the PSF at 850$\mu$m. 
When integrating the extended emission by sky-subtracted aperture photometry, 
the net flux (as listed in Table 2)
is 756 mJy at 450$\mu$m and 271 mJy
at 850$\mu$m.
The sub-millimeter source
coincides with a faint, slightly extended source on our {\it K}-band
image (Fig. 5, bottom panel)
at position 03:28:34.5 +31:07:05 (J2000). The source, which
we will refer to as 
NGC1333 J032834.5+310705 (K)
is also faintly 
visible on the 2MASS {\it K}-band image. It can be seen in our S(1) image, 
albeit significantly
fainter, indicating a very red color. It is invisible
on the optical CCD images by \citet{bal96} and their later CCD data.
NGC1333 J032834.5+310705 (K)
has a cometary morphology, with the 
apex of the conical shape pointing to the east. On the basis of
the strong compact sub-millimeter flux, and its near-infrared 
detectability and morphology we tentatively identify this source as a class I
object. It cannot be determined if this object is physically 
close to the IRAS 03256+3055
group of sub-millimeter clumps or whether
it is farther away along the line of sight.

\section{Summary}

We have presented optical and near-infrared images of the
system of Herbig-Haro objects 340 B, 343 A-F, and their
driving source, 
NGC1333 J032845.3+310542 (K)
in NGC~1333.
The Herbig-Haro objects and the H$_2$ emission seen in the {\it K} band
outline a jet of centro-symmetric S-shaped morphology,
suggesting a precessing driving source.
At wavelengths shorter than 2.2$\mu$m, the central driving
source is dominated by line emission from the south-western part
of the jet, which is moving toward the observer.
In the {\it K}-band, the flux is dominated by continuum radiation, 
polarized and therefore partly scattered, from the central source. 
The spectrum
of this source is a steep continuum with some Br$\gamma$ emission,
characteristic of a class I source.
The SED of the region based on sub-millimeter and IRAS data, suggests 
an evolutionary state of class 0 or I.
The sub-millimeter maps at 450 and 850 $\mu$m confirm the existence of a 
region of extended emission, probably a cluster of pre-stellar
clumps. Clump D associated with 
the near-infrared object NGC1333 J032845.3+310542 (K)
is probably the first
in this cluster to begin forming a star.

\acknowledgements

The United Kingdom Infrared Telescope is operated by 
the Joint Astronomy Centre on behalf of the 
U.K. Particle Physics and Astronomy Research Council. 
We thank the Department of Physical Sciences, 
University of Hertfordshire for providing IRPOL2 for the UKIRT.

The James Clerk Maxwell Telescope is operated by 
The Joint Astronomy Centre on behalf of the 
Particle Physics and Astronomy Research Council of the United Kingdom, 
the Netherlands Organisation for Scientific Research, 
and the National Research Council of Canada. 

This paper is based in part on data collected at the Subaru Telescope, 
which is operated by the National Astronomical Observatory of Japan. 

This publication makes use of data products from the 
Two Micron All Sky Survey, which is a joint project 
of the University of Massachusetts and the 
Infrared Processing and Analysis Center / California 
Institute of Technology, funded by the 
National Aeronautics and Space Administration 
and the National Science Foundation.

NOAO is operated by the Association of Universities for Research in Astronomy (AURA),
Inc. under cooperative agreement with the National Science Foundation.

\end{document}